\documentclass[english,aip,reprint]{revtex4-1}
\pdfoutput=1
\usepackage{graphicx}
\usepackage{bm}
\usepackage{color}
\usepackage{natbib}

\begin{document}

\title{\textcolor{blue}{Measurement of the Transmission Phase of an Electron in a Quantum Two-Path Interferometer}}

\author{S.~\surname{Takada}}
\email{shintaro.takada@neel.cnrs.fr}
\affiliation{Department of Applied Physics, University of Tokyo, Bunkyo-ku, Tokyo, 113-8656, Japan}

\author{M.~\surname{Yamamoto}}
\affiliation{Department of Applied Physics, University of Tokyo, Bunkyo-ku, Tokyo, 113-8656, Japan}
\affiliation{PRESTO, JST, Kawaguchi-shi, Saitama 331-0012, Japan}

\author{C.~\surname{B\"{a}uerle}}
\affiliation{Univ. Grenoble Alpes, Institut NEEL, F-38042 Grenoble, France}
\affiliation{CNRS, Institut NEEL, F-38042 Grenoble, France}

\author{K.~\surname{Watanabe}}
\affiliation{Department of Applied Physics, University of Tokyo, Bunkyo-ku, Tokyo, 113-8656, Japan}

\author{A.~\surname{Ludwig}}
\affiliation{Lehrstuhl f\"{u}r Angewandte Festk\"{o}rperphysik, Ruhr-Universit\"{a}t Bochum, Universit\"{a}tsstra\ss e 150, 44780 Bochum, Germany}

\author{A. D.~\surname{Wieck}}
\affiliation{Lehrstuhl f\"{u}r Angewandte Festk\"{o}rperphysik, Ruhr-Universit\"{a}t Bochum, Universit\"{a}tsstra\ss e 150, 44780 Bochum, Germany}

\author{S.~\surname{Tarucha}}
\affiliation{Department of Applied Physics, University of Tokyo, Bunkyo-ku, Tokyo, 113-8656, Japan}
\affiliation{RIKEN, Center for Emergent Matter Science (CEMS), Wako, Saitama, 351-0198, Japan}

\date{\today}

\begin{abstract}
A quantum two-path interferometer allows for direct measurement of the transmission phase shift of an electron, providing useful information on coherent scattering problems.
In mesoscopic systems, however, the two-path interference is easily smeared by contributions from other paths, and this makes it difficult to observe the \textit{true} transmission phase shift.
To eliminate this problem, multi-terminal Aharonov-Bohm (AB) interferometers have been used to derive the phase shift by assuming that the relative phase shift of the electrons between the two paths is simply obtained when a smooth shift of the AB oscillations is observed.
Nevertheless the phase shifts using such a criterion have sometimes been inconsistent with theory.
On the other hand, we have used an AB ring contacted to tunnel-coupled wires and acquired the phase shift consistent with theory when the two output currents through the coupled wires oscillate with well-defined anti-phase.
Here, we investigate thoroughly these two criteria used to ensure a reliable phase measurement, the anti-phase relation of the two output currents and the smooth phase shift in the AB oscillation.
We confirm that the well-defined anti-phase relation ensures a correct phase measurement with a quantum two-path interference.
In contrast we find that even in a situation where the anti-phase relation is less well-defined, the smooth phase shift in the AB oscillation can still occur but does not give the correct transmission phase due to contributions from multiple paths.
This indicates that the phase relation of the two output currents in our interferometer gives a good criterion for the measurement of the \textit{true} transmission phase while the smooth phase shift in the AB oscillation itself does not.

\end{abstract}

\maketitle
The transmission phase of an electron plays a crucial role in various quantum interference phenomena.
Full characterization of the coherent transport therefore requires a reliable phase measurement, but this is still challenging.
One may envisage a quantum two-path interferometer because the interference is measured as a function of the phase difference between the two paths.
For instance the phase shift across a quantum dot (QD), in which one can control the quantum state of single electrons, can be measured using a QD embedded in one of the two arms of the interferometer.
The theory predicts a Breit-Wigner type phase shift across a Coulomb peak (CP) \cite{Yeyati1995} and a $\pi/2$ phase shift across a Kondo-singlet state \cite{Gerland2000} and both were experimentally investigated.

The Breit-Wigner type phase shift was confirmed by a pioneering experiment for a QD embedded in a multi-terminal Aharonov-Bohm (AB) interferometer \cite{Schuster1997}.
The phase shift was derived from a smooth shift of AB oscillation phase.
However, unanticipated results have sometimes been observed, such as a universal phase lapse between CPs for a large QD \cite{Schuster1997, Heiblum_phase_2005} and a large phase shift exceeding $\pi$ across two Coulomb peaks of a spin degenerated level for a Kondo correlated QD \cite{Ji2000, Ji2002}.
Although several mechanisms have been proposed to account for the universal phase lapse \cite{Baltin1999, Silvestrov2000, Yeyati2000, Hackenbroich2001, Karrasch2007, Molina2012}, origins of the behavior remain unaccounted.
This is also related to the fact that only a few experiments have been reported for the phase measurement \cite{Schuster1997, Heiblum_phase_2005, Aikawa2004B} due to difficulty in realizing a reliable phase measurement for QDs.
In a two-terminal AB interferometer, which is usually considered as a two-path interferometer, the phase of the AB oscillation is fixed to either $0$ or $\pi$ at zero magnetic field due to the boundary conditions imposed by the two-terminal geometry, whereas the real transmission phase across the QD is not.
The $0$\,-$\pi$ rigidity of the observed phase called phase rigidity \cite{Yeyati1995, Yacoby1996} therefore implies that the two-terminal AB ring is not a \textit{true} two-path interferometer; because not only direct two paths but also paths of an electron encircling the AB ring multiple times contribute to the interference.

A multi-terminal \cite{Schuster1997, Ji2000, Ji2002, Heiblum_phase_2005, Sigrist2004, Sven2010} as well as a multi-channel \cite{Katsumoto2006} AB interferometer was employed to avoid the phase rigidity and to measure the transmission phase shift across a gate-defined QD embedded in one of the two arms.
In these experiments lifting of the phase rigidity was confirmed by observation of a smooth phase shift with gate voltage at a fixed magnetic field.
On the other hand lifting of the phase rigidity does not readily ensure that the observed interference is a \textit{pure} two-path interference.
There is a possibility that contributions from multi-path interferences \cite{Entin2002, Simon2005} still remain.
Previously we have developed a new type of interferometer realized in an AB ring contacted to tunnel-coupled wires.
It can be tuned into a two-path interferometer in the weak tunnel-coupling regime when the two output currents through the two coupled wires oscillate with magnetic field but opposite phase \cite{Yamamoto:2012fk, Tobias2014, Aharony2014}.
We have used this original device to investigate the transmission phase shift across a Kondo correlated QD and obtained a very good agreement for the phase shift between experiment and theory by carefully analyzing the anti-phase oscillations \cite{Takada2014}.
In addition we have noticed that a smooth phase shift as a function of gate voltage can be observed even when the contributions from other than the direct two-paths exist.
Here a question, on how reliable the phase measurement in such a situation is, is raised.
This is indeed a serious problem because all previous experiments relied on the observation of such a smooth phase shift to derive the phase shift and the results often showed disagreement with theory.

In this letter, we experimentally address this question.
We investigate the influence of multi-path interferences on the phase measurement by analyzing both anti-phase and non-anti-phase AB oscillations between the two output currents through the coupled wires.
We show that the smooth phase shift at a fixed magnetic field is observed even when contributions of interferences from multiple paths are present.
In this case, however, we observe no well-defined anti-phase AB oscillations and find that the measured phase shift deviates significantly from the theoretically expected transmission phase shift.
In contrast when we observe the anti-phase AB oscillation, the derived phase shift is in very good agreement with theory.
We thus conclude that the anti-phase oscillations of the two output currents are a hallmark of a reliable phase measurement while the smooth phase shift as observed for a multi-terminal geometry is not.

%Fig. 1 (Device A) %%%%%%%%%%%%%%%%%%%%%%
\begin{figure}[t]
\centering
\includegraphics*[width=1.00\columnwidth]{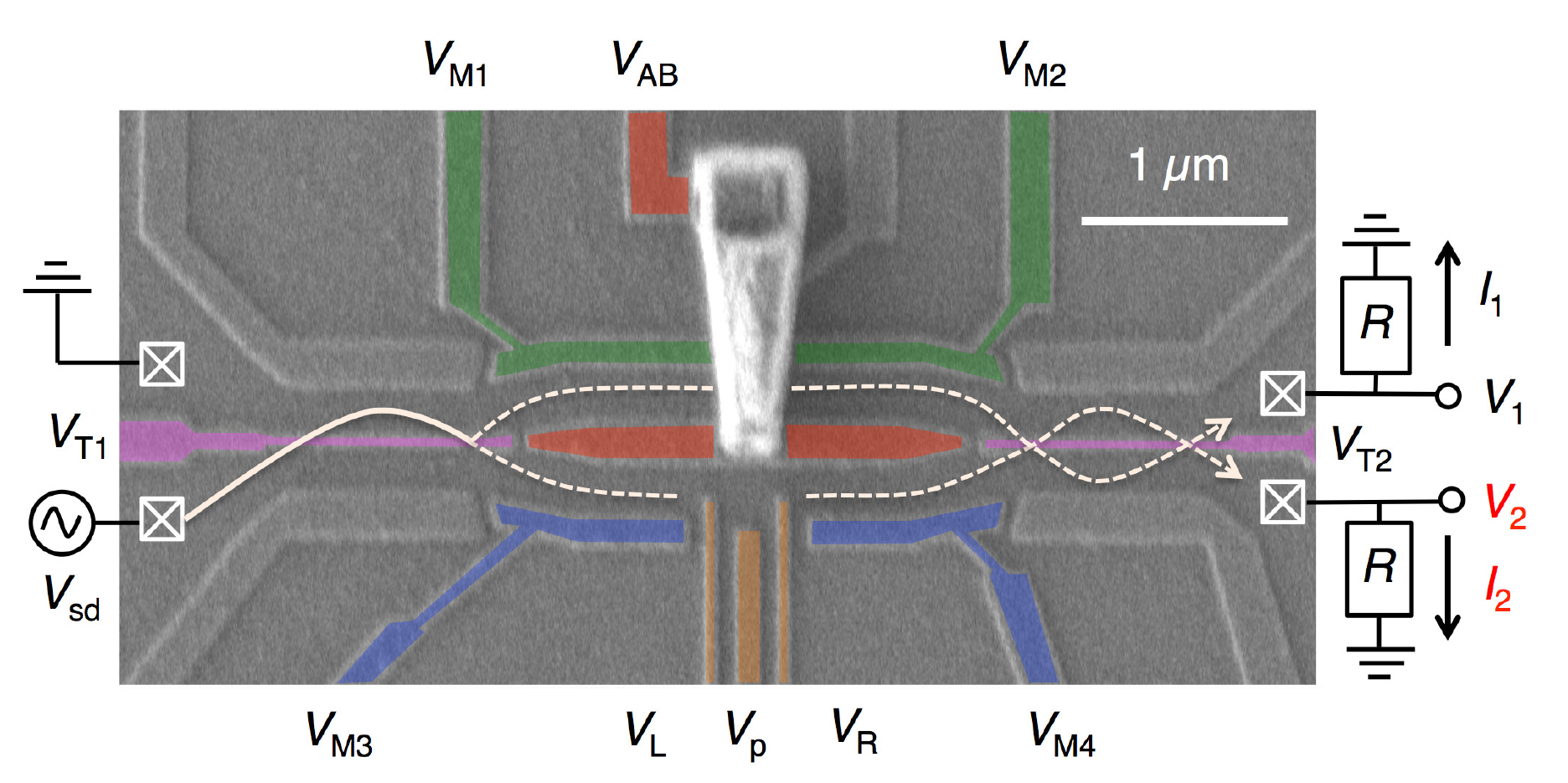}
\caption{\label{DeviceA} SEM picture of device A and measurement setup. Output currents are measured for a constant voltage bias across the resistance $R = 10 ~{\rm k\Omega}$. Dashed lines indicate electron trajectories for the two-path interference.}
\end{figure}
\begin{figure}[t]
\centering
\includegraphics*[width=1.00\columnwidth]{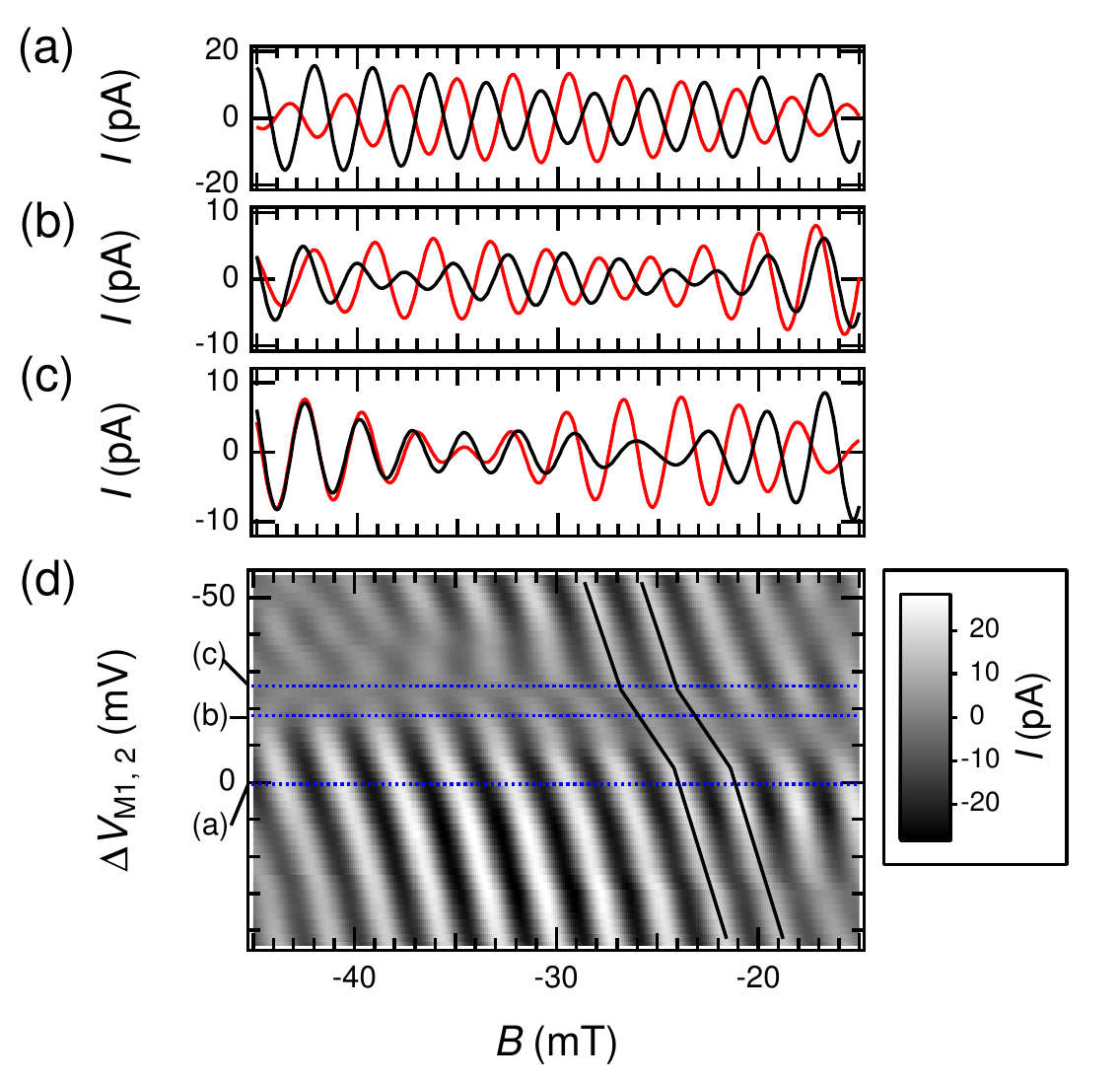}
\caption{\label{PhaseShift} (a), (b), (c) Quantum oscillations as a function of magnetic field $B$ observed in $I_{\rm 1}$ (black line) and $I_{\rm 2}$ (red line) in the weak tunnel-coupling regime. Only oscillating parts extracted from raw data by performing a complex fast Fourier transform (FFT) are plotted. Three figures are measured at the different gate voltages of $\Delta V_{\rm M1, 2}$, which are indicated in (d). (d) Modulation of geometrical phase as a function of $B$ and $\Delta V_{\rm M1, 2}$. The oscillating components with $B$ extracted from a complex FFT of ($I=I_{\rm 1} - I_{\rm 2}$) are plotted in the plane of $B$ and $\Delta V_{\rm M1, 2}$. The black solid lines are added to highlight the change of the slope.}
\end{figure}
\begin{figure}[t]
\centering
\includegraphics*[width=\columnwidth]{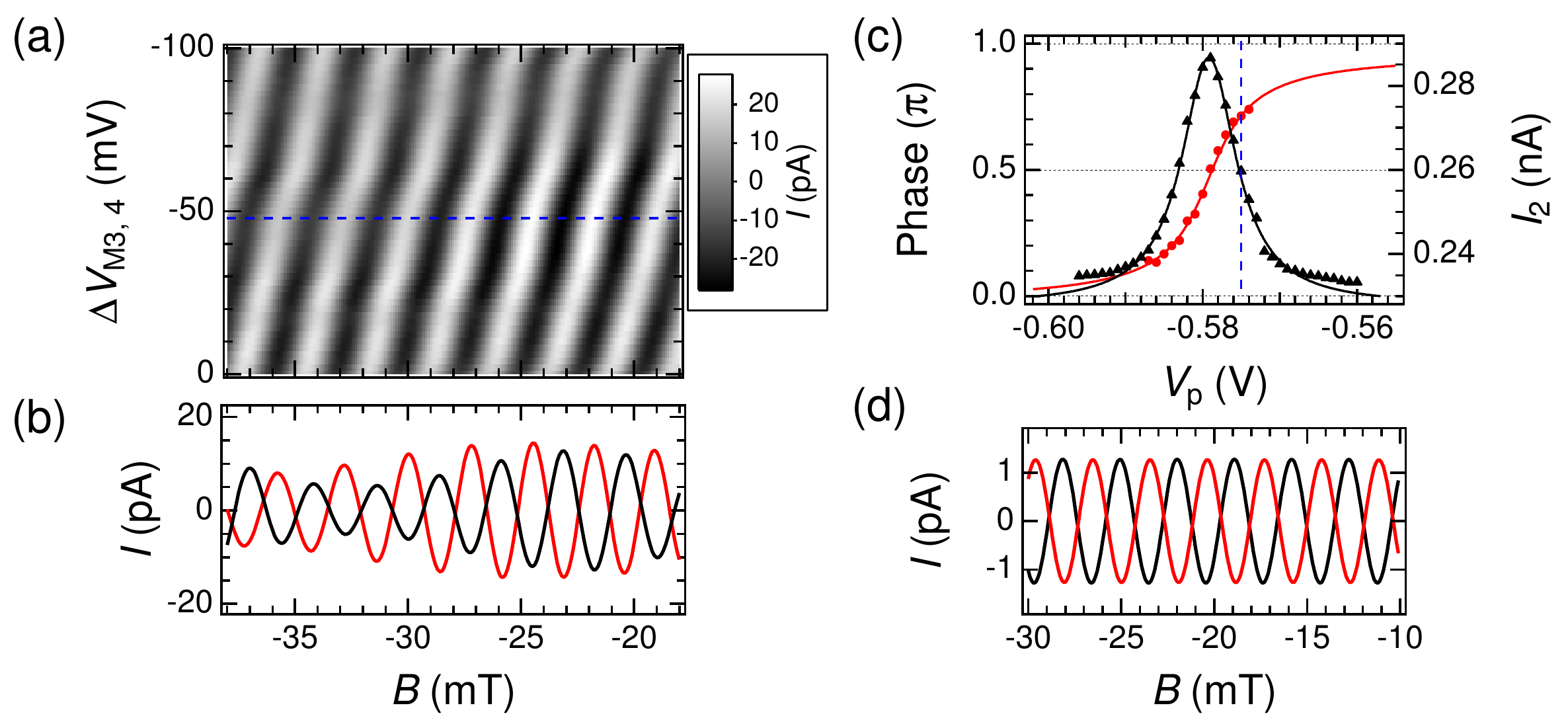}
\caption{\label{APShift}(a) Transmission phase shift by modulation of geometrical phase in anti-phase configuration. Quantum oscillations as a function of magnetic field extracted from the FFT analysis of ($I=I_{\rm 1} - I_{\rm 2}$) are plotted for different gate voltage shifts of $\Delta V_{\rm M3, 4}$. (b) The oscillating part of $I_{\rm 1}$ (black) and $I_{\rm 2}$ (red) of the data shown in (a) at $\Delta V_{\rm M3, 4}$ (blue dashed line). (c) Transmission phase shift across a Coulomb peak in the anti-phase configuration. The phase obtained from experiment is shown by the red circles for left axis with the phase behavior expected theoretically (red solid line). The $I_{\rm 2}$ averaged over one oscillation period of magnetic field is plotted on the right axis with the Lorentzian fit of $I_{\rm 2}$ (black solid line). (d) Oscillating part of $I_{\rm 1}$ (black) and $I_{\rm 2}$ (red) of the data shown in (c) at $V_{\rm p}$ indicated by the blue dashed line.}
\end{figure}
\begin{figure}[t]
\centering
\includegraphics*[width=0.90\columnwidth]{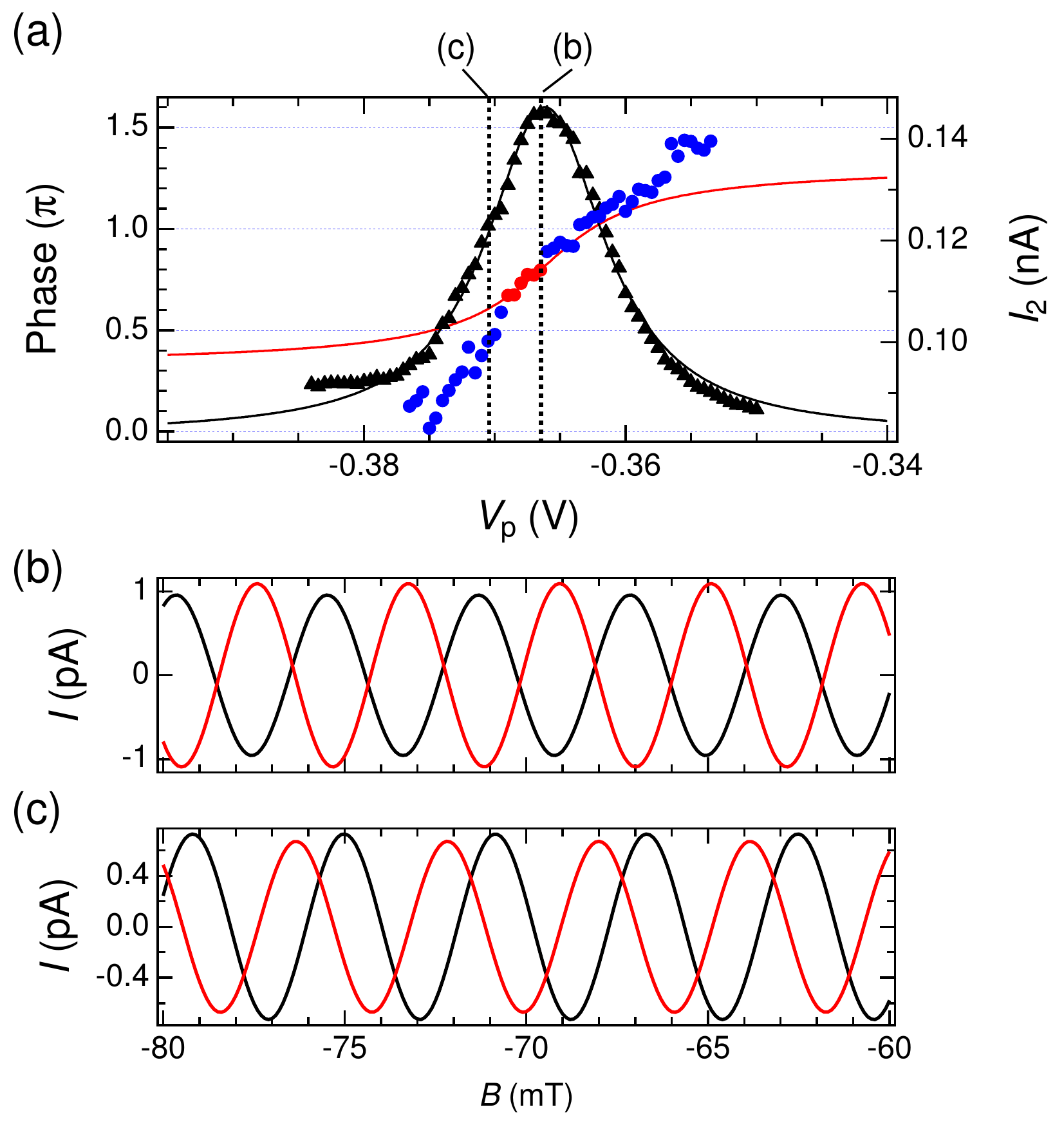}
\caption{\label{IPShift}(a) Influence of multi-path interference on the transmission phase shift across a Coulomb peak. The phase shift indicated by the blue (red) circles are extracted from ($I_{\rm 1} - I_{\rm 2}$) for the oscillating two output currents with poorly (well) defined anti-phase. The red solid line shows the calculation of the phase using Friedel's sum rule. The black triangles indicate the measured current $I_{\rm 2}$ at each $V_{\rm p}$ and the black solid line is the Lorentzian fit. (b), (c) Quantum oscillations of the two output currents $I_{\rm 1}$ in black and $I_{\rm 2}$ in red measured at the $V_{\rm p}$ indicated by (b) and (c) in (a). Oscillating components extracted by a complex FFT are plotted here.}
\end{figure}

%%%%%%%%%%%%%%%%%%%%%%%
The device was fabricated on a two-dimensional electron gas formed in a GaAs/AlGaAs heterostructure [electron density $n = 3.21 \times 10^{11}~{\rm cm}^{-2}$, electron mobility $\mu = 8.6 \times 10^5 {\rm cm}^{2}/{\rm Vs}$ at the temperature of $T = 4.2$ K; see Fig.\,\ref{DeviceA}].
The interferometer was defined by applying negative voltages on surface Schottky gates and locally depleting electrons underneath the gates.
It consists of an AB ring at the center and tunnel-coupled wires on both ends of the ring.
The coupling energy of the tunnel-coupled wires can be controlled by the gate voltages $V_{\rm T1}$ and $V_{\rm T2}$.
The gate voltages $V_{\rm M1}$, $V_{\rm M2}$ ($V_{\rm M3}$, $V_{\rm M4}$) are used to modulate the wave vector of electrons in the upper (lower) path.
A QD can also be formed by applying the gate voltages $V_{\rm L}$, $V_{\rm p}$ and $V_{\rm R}$.
We measured two samples with a slightly different size of the AB ring and QD (device A and B).
The data shown in Fig.\,\ref{PhaseShift} and Fig.\,\ref{APShift} was measured in device A and that in Fig.\,\ref{IPShift} for device B.
Electrons are injected from the lower left contact by applying an AC bias ($20 \sim 100\,{\rm \mu V}$, 23.3 Hz) and currents are measured at the two right contacts by voltage measurements across the resistance ($I_{\rm 1(2)} = V_{\rm 1(2)} / R$) using a standard lock-in technique.

%Fig. 2 (Device A) %%%%%%%%%%%%%%%%%%%%%%
%%%%%%%%%%%%%%%%%%%%%%%
We first tuned the tunnel-coupled wires into the weak coupling regime such that the interferometer works as a two-path interferometer, where the two output currents oscillate with anti-phase as shown in Fig.\,\ref{PhaseShift}(a).
For panels (a) - (c) of Fig.\,\ref{PhaseShift} we plot the oscillating components of the currents as a function of magnetic field, which are obtained by performing a complex fast Fourier transform (FFT) of the raw data, filtering out the noise outside the oscillation frequency and performing a back transform.
The two-path interference is sensitive to the difference of the transmission phase shift between the two paths across the AB ring $\theta = \oint {\bm k}\cdot {\rm d}{\bm l} - \frac{e}{\hbar}BS + \varphi_{\rm dot}$.
The first term is the geometrical phase depending on the path length ${\bm l}$ and the wave vector of an electron ${\bm k}$, the second term is the AB phase controlled by the magnetic field $B$ penetrating the surface area $S$ enclosed by the two paths, and the third term is the transmission phase shift across the QD, respectively.
Fig.\,\ref{PhaseShift}(a) shows the phase shift induced by the modulation of the AB phase.

We then measured the phase shift induced by modulation of the geometrical phase, where the wave vector of electrons passing through the upper path is controlled by the gate voltages $V_{\rm M1}$ and $V_{\rm M2}$.
Here $V_{\rm M1}$ and $V_{\rm M2}$ are shifted simultaneously by the same amount.
The result is shown in Fig.\,\ref{PhaseShift}(d).
The oscillating part of $I=I_{\rm 1}-I_{\rm 2}$ as a function of magnetic field, which mainly consists of the anti-phase components, is plotted for the gate voltage shift $V_{\rm M1, 2}$ along the vertical axis around the configuration used for the measurement of Fig.\,\ref{PhaseShift}(a).
Around $\Delta V_{\rm M1, 2}=0$, where the anti-phase oscillations of the two output currents are observed, the phase smoothly shifts along the vertical axis with a certain slope.
Around the gate voltage shift from $-5$ mV to $-25$ mV and the magnetic field range from $-15$ mT to $-30$ mT, the phase smoothly shifts as well but with a slightly different slope as indicated with the black solid lines, where the two output currents do not oscillate with anti-phase as shown in Fig.\,\ref{PhaseShift}(b).
For the more negative voltage shift and the magnetic field range from $-30$ mT to $-45$ mT, abrupt phase jumps of $\pi$ along the vertical axis are observed similarly to a two-terminal device that suffers from the phase rigidity.
In this region the two output currents oscillate in phase as shown in Fig.\,\ref{PhaseShift}(c).

The anti-phase oscillations of the two output currents indicate that the total current ($I_{\rm 1} + I_{\rm 2}$) is independent on $\theta$.
This is a clear indication that interferences coming from encircling paths around the AB ring are absent and hence the realization of the \textit{pure} two-path interference as depicted by the dashed lines in Fig.\,\ref{DeviceA}.
On the other hand, when the two output currents do not oscillate with anti-phase, paths encircling the AB ring also contribute to the interference even though the magneto oscillations still show a smooth phase shift as a function of gate voltages at a fixed magnetic field.
In such a case, however, the observed phase shift is modified from the \textit{true} transmission phase shift as we will demonstrate in the following.

%Fig. 3 (Device A) %%%%%%%%%%%%%
%%%%%%%%%%%%%%%%%%%%%%%%%%%%%%%%%%%%%%%
First we show that the phase relation between the two output currents is a good criteria to exclude the contributions of multi-path interferences and allows for a reliable measurement of the transmission phase shift.
For this we carefully tuned the interferometer to observe the anti-phase oscillations as shown in Fig.\,\ref{APShift}(b).
For this condition, we observed a smooth phase shift induced by the modulation of the geometrical phase through $V_{\rm M3, 4}$ with a single constant slope [Fig.\,\ref{APShift}(a)].
At the same time we also measure the transmission phase shift across a QD, where the experimental results can be compared with theory \cite{Yeyati1995, Schuster1997, Takada2014} [Fig.\,\ref{APShift}(c) and (d)].
The QD is formed by tuning the gate voltages  $V_{\rm L}$, $V_{\rm p}$ and $V_{\rm R}$ and the phase shift across a CP is observed by recording quantum oscillations as a function of magnetic field at each value of the plunger gate voltage $V_{\rm p}$.
This result is presented in Fig.\,\ref{APShift}(c).
The current $I_{\rm 2}$ averaged over one oscillation period of the magnetic field mimics the shape of the CP with a finite background current coming from the current through the upper path of the AB ring.
The black solid line is a Lorentzian fit of the CP, which is used to calculate the transmission phase shift expected from Friedel's sum rule and depicted by the red solid line \cite{Yeyati1995, Schuster1997, Takada2014}.
The numerical values of the observed phase shift are obtained from a complex FFT of ($I_{\rm 1} - I_{\rm 2}$).
The observed phase shift is in good agreement with the theoretically expected $\pi$-phase shift.
This result confirms that the phase evolution obtained under the condition of anti-phase oscillations of the two output currents is the \textit{true} transmission phase shift observed for the \textit{pure} two-path interference.

%Fig. 4 (Device B) %%%%%%%%%%%%%%%%%%
%%%%%%%%%%%%%%%%%%%%%%%%%%%%%%%%%%%%%%%%%%%%%%%%%
We now turn to the phase shift measurements when the two output currents are not kept anti-phase over the entire gate voltage ($V_{\rm p}$) scan across a CP.
The measured phase shift is shown in Fig.\,\ref{IPShift}(a).
The phase smoothly shifts across the CP by $1.5\pi$, which is inconsistent with the $\pi$-phase shift expected from Friedel's sum rule (red solid line).
In this data the two output currents oscillate with anti-phase for $V_{\rm p}$ only around the center of the CP (red circles) as shown in Fig.\,\ref{IPShift}(b).
For the entire other range (blue circles) they do not oscillate with anti-phase as shown in Fig.\,\ref{IPShift}(c) and hence the measured phase shift must contain contributions from multi-path interferences.
The larger phase shift observed here must therefore come from the additional multi-path contributions.
Such contributions from multi-path interferences might explain the unexpected large phase shift across Kondo correlated Coulomb peaks observed in the previous experiments \cite{Ji2000, Ji2002}.
Note that we consider the oscillations as non-anti-phase when the phase difference between the two outputs is deviating more than $10\%$ ($\sim 0.2\pi$) from the anti-phase.
The phase measurements with anti-phase oscillations within this error are in good agreement with theoretical expectations as shown in Fig.\,\ref{APShift}(c).

In the weak tunnel-coupling regime the device has four terminals and hence each output current is not bound to the phase rigidity.
This allows for observation of a smooth phase shift induced by the gate voltage at a fixed magnetic field.
However in case we fail to keep an anti-phase relation between the two output currents, the obtained phase shift can be modified by multi-path contributions and the phase shift is inconsistent with theory.

Finally we discuss the key to realize a \textit{pure} two-path interference in an AB ring contacted to tunnel-coupled wires.
As we already pointed out in our earlier experimental \cite{Yamamoto:2012fk} and theoretical \cite{Tobias2014, Aharony2014} works, the most important factor is to make the tunnel-coupling weak enough to suppress the encircling paths.
In addition a smooth potential connection between the AB ring and the tunnel-coupled wires is important.
As seen from Fig.\,\ref{PhaseShift}(d) the gate voltages $V_{\rm M1}$ and $V_{\rm M2}$ play a crucial role to realize the anti-phase oscillations or two-path interference.
The gate voltage $V_{\rm M1}$ and $V_{\rm M2}$ are not effective for the tunnel-coupling strength but effective for the potential profile at the transition regions between the ring and the coupled wires.
This suggests that the key is not only the weak tunnel-coupling but also a smooth potential connection between the AB ring and the tunnel-coupled wires.
In other words, one needs to suppress backscattering of an electron into the other path at this transition region.
Indeed the importance of the smooth potential connection is also mentioned in ref.~\onlinecite{Tobias2014}.
However, note that ``smooth” here is not with respect to the Fermi wave-length: since the 2DEG is $100$~nm away from the gate electrodes, the potential profile is smooth with respect to the Fermi wavelength for all gate voltages in Fig.\,\ref{PhaseShift}(d).
The required smoothness depends on the tunnel-coupling energy and the potential profile of the two wires at the transition regions, although it is difficult to explore experimentally the detail of the connection of the wave function due to the existence of many channels in each path.

In summary, we employed an AB ring with tunnel-coupled wires to demonstrate how to measure the \textit{true} transmission phase of an electron.
We find that lifting the phase rigidity, i.e., the observation of a smooth phase shift at a fixed magnetic field in a multi-terminal AB interferometer does not ensure a correct measurement of the \textit{true} transmission phase.
Our original AB interferometer, on the contrary, allows for the measurement of the \textit{true} transmission phase shift by simply tuning it into a regime where the two output currents oscillate anti-phase.
This interferometer is hence extremely suitable to investigate unsolved problems related to the transmission phase such as the universal phase behavior for large quantum dots \cite{Schuster1997, Heiblum_phase_2005}.

S. Takada acknowledges financial support from JSPS Research Fellowships for Young Scientists, French Government Scholarship for Scientific Disciplines and the European Union’s Horizon 2020 research and innovation program under the Marie Sklodowska-Curie grant agreement No 654603. M.Y. acknowledges financial support by Grant-in-Aid for Young Scientists A (No. 23684019) and Grant-in-Aid for Challenging Exploratory Research (No. 25610070). C. B. acknowledges financial support from the French National Agency (ANR) in the frame of its program BLANC FLYELEC Project No. anr-12BS10-001, as well as from DRECI-CNRS/JSPS (PRC 0677) international collaboration. A.L. and A.D.W. acknowledge gratefully support of Mercur  Pr-2013-0001, DFG-TRR160,  BMBF - Q.com-H  16KIS0109, and the DFH/UFA  CDFA-05-06. S. Tarucha acknowledges financial support by JSPS, Grant-in-Aid for Scientific Research S (No. 26220710), MEXT project for Developing Innovation Systems, and JST Strategic International Cooperative Program.

\end{document}